\documentstyle[12pt,psfig,aaspp4]{article}

\newcommand{\solarm}{{\rm M_\odot}}
\newcommand{\kira}{\mbox {${\sf Kira}$}}
\newcommand{\seba}{\mbox {${\sf SeBa}$}}
\newcommand{\msun}{\mbox {$M_\odot$}}

\lefthead{K.\ Takahashi \& S.\ F.\ Portegies Zwart}
\righthead{Disruption of globular clusters}

\begin{document}

\title{
The disruption of globular star clusters in the galaxy: \\
A comparative analysis between Fokker-Planck and $N$-body models.}

\author{
Koji Takahashi
and Simon F. Portegies Zwart$^\star$
}
\affil{Department of Earth Sciences and Astronomy,
       College of Arts and Sciences, University of Tokyo,
       \it3-8-1 Komaba, Meguro-ku, Tokyo 153-8902, Japan
}

$^\star$ Japan Society for the Promotion of Science Fellow \\

\authoremail{takahasi@grape.c.u-tokyo.ac.jp, spz@grape.c.u-tokyo.ac.jp}

\date{}


\clearpage

\begin{abstract}
Recent $N$-body simulations have shown that there is a serious
discrepancy between the results of the $N$-body simulations and the
results of Fokker-Planck simulations for the evolution of globular and
rich open clusters under the influence of the galactic tidal field.
In some cases, the lifetime obtained by Fokker-Planck calculations is
more than an order of magnitude smaller than those by $N$-body
simulations.  In this letter we show that the principal cause for this
discrepancy is an over-simplified treatment of the tidal field used in
previous Fokker-Planck simulations.  We performed new Fokker-Planck
calculations using a more appropriate implementation of the boundary
condition of the tidal field.  The implementation is only possible
with {\it anisotropic} Fokker-Planck models, while all previous
Fokker-Planck calculations rely on the assumption of isotropy.  Our
new Fokker-Planck results agree well with $N$-body results.
Comparison of the two types of simulations gives a better
understanding of the cluster evolution.
\end{abstract}



\section{Introduction}
Star clusters range in mass from a few hundred to several million
solar-masses.  In order to understand their formation and dynamical
evolution, detailed numerical modeling is required. There are,
however, many effects which complicate their evolution and numerical
models of star clusters are just beginning to incorporate deviations
from the ideal star cluster (see Vesperini \& Heggie 1997; Portegies
Zwart et al.\ 1998a).  \nocite{vh97}\nocite{pztl98}

Collisional $N$-body simulations are very expensive in terms of
computer time. Even with supercomputers or special-purpose
machines, it is impossible to do a simulation with the number
of particles comparable to that of a real globular cluster. Therefore 
we are forced to rely on either $N$-body simulations with smaller
number of particles or more approximate methods such as Fokker-Planck
techniques. In theory, these two approaches should give identical
results. 

In order to check the reliability of the Fokker-Planck models with
other models ($N$-body, gaseous, Monte-Carlo, etc.), some authors
compared the results of various types of numerical simulations
(Aarseth et al.\ 1974; Giersz and Heggie 1994a, 1994b; Giersz and
Spurzem 1994; Spurzem and Takahashi 1995). These comparisons
demonstrate that for isolated clusters made of point masses
the results of Fokker-Planck simulations are in good agreement with
$N$-body computations.

Recent comparison between the same techniques for clusters in the
galactic tidal field, however, gave a completely different view
(Fukushige \& Heggie 1995; Heggie et al.\@ 1998);
the result of the $N$-body simulations did not seem to converge to 
that of the Fokker-Planck simulations in the limit for $N \rightarrow \infty$,
contrary to what was expected.

The disagreement between Fokker-Planck models and $N$-body models was
even more clearly shown by Portegies Zwart et al.\@ (1998b, PZHMM).
They performed a series of $N$-body simulations with up to 32768 stars
with identical initial conditions as one of the Fokker-Planck
simulations of Chernoff and Weinberg (1990, CW).

The results of the computations of PZHMM can be summarized as follows:
1) The $N$-body model with the largest number of particles has a
lifetime more than an order of magnitude longer than that of the
comparable model of CW.  2) The lifetime of the cluster 
depends on the number of stars in a rather complex way.
Since the fundamental assumption of Fokker-Planck calculations is that 
the evolution is independent of the number of stars, the results of
PZHMM might imply that the results of Fokker-Planck calculations for
clusters in a tidal field and with stellar evolution are of
questionable validity.

The purpose of this letter is to explore what caused this discrepancy
between the $N$-body models of PZHMM and the Fokker-Planck models
of CW.

\section{The Models}

\subsection{The $N$-body model}
The direct $N$-body integration program \kira\ (Hut 1994; Hut et al.\
1995) is used in combination with the stellar evolution package \seba\
(Portegies Zwart \& Verbunt 1996; Portegies Zwart \& Yungelson 1998).
\nocite{hut94}\nocite{hmm95}\nocite{pzv96}\nocite{pzy98} Both models
are part of the Starlab software tool set (version 3.1, for the
details of its implementation see PZHMM).\nocite{pzhmm98}

The numerical integration of the motion of the stars is performed
using a fourth-order individual--time-step Hermite scheme (Makino and
Aarseth 1992).\nocite{ma92} For all $N$-body
simulations we used GRAPE-4 (Makino et al. 1997).

\subsection{The Fokker-Planck model}

The model used by CW is an orbit-averaged Fokker-Planck scheme in
which the velocity distribution of the stars is assumed to be {\it
isotropic}.  In this paper we report the results of an {\em
anisotropic} Fokker-Planck scheme in which the distribution function
$f$ depends both on energy $E$ and angular momentum $J$. The
two-dimensional orbit-averaged Fokker-Planck equation in $(E,J)$-space
is solved numerically (see Cohn 1979; Takahashi 1995, 1997; Takahashi
et al\@. 1997).  Although anisotropy is usually unimportant in the
central parts of the clusters, it is significant in the outer parts.
Therefore we expect that the effects of anisotropy on the escape rate
of stars from the clusters can be large.  Furthermore we have to
consider $J$-dependence of the distribution function when we like to
use a realistic escape criterion (see below).

\begin{figure}
\psfig{file=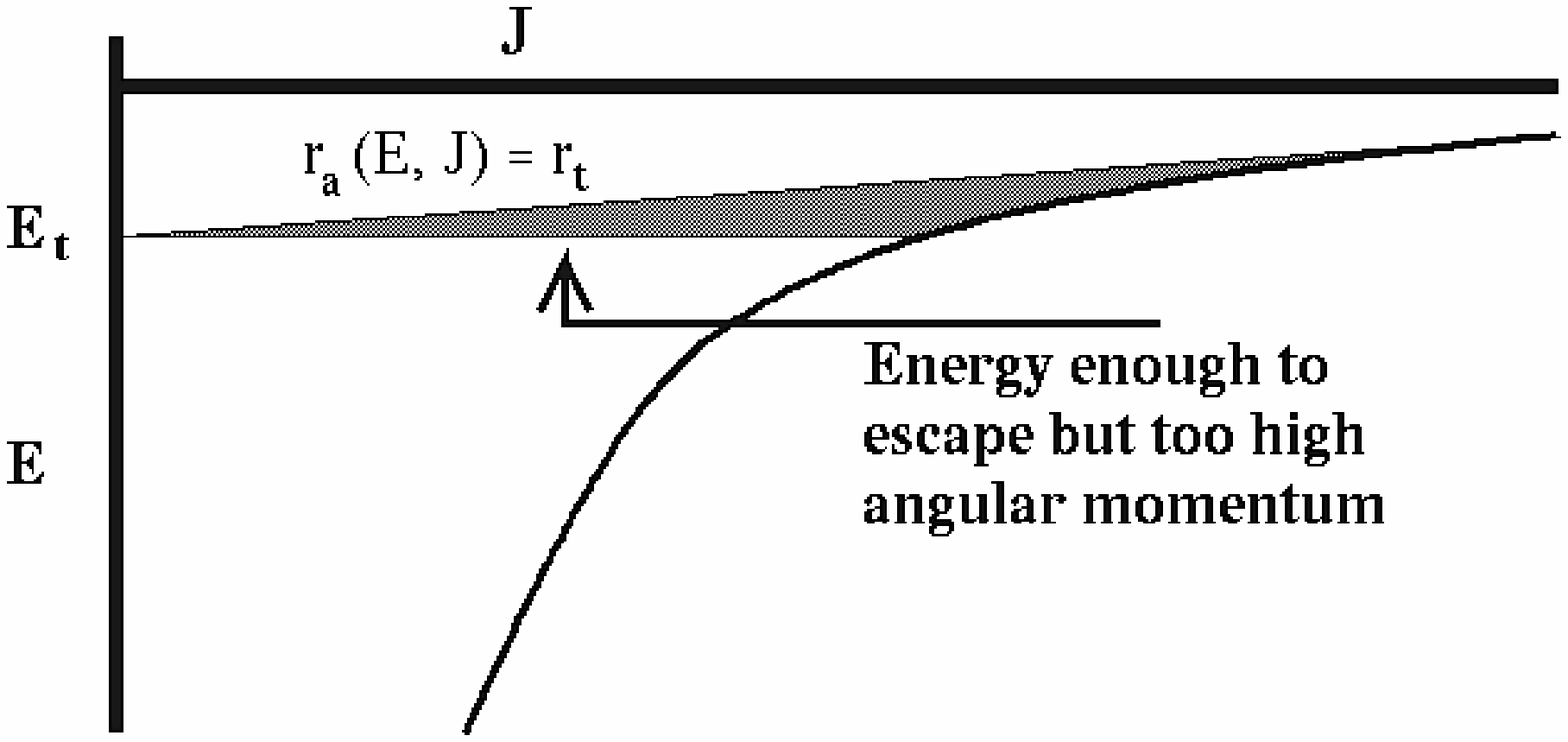,bbllx=30pt,bblly=40pt,bburx=570pt,bbury=300pt,width=10.0cm}
\figcaption{ Schematic diagram of the energy--angular momentum plane
for a star cluster. Using the energy criterion all stars with energy
$E$ greater than the energy at the tidal radius $E_{\rm t}$ escape. 
If the apocenter criterion is applied paricles in the schaded region
are not in an escape orbit and therefore remain bound to the cluster.
\label{fig:Pot}
}
\end{figure}

In CW's isotropic model, a star is removed from the stellar system
when its energy exceeds the potential energy at the tidal radius
$r_{\rm t}$ (which we will call the {\em energy criterion}). In an
isotropic Fokker-Planck model, one has no choice but to use the energy
as a criterion for escape. In the anisotropic model a more realistic
escape condition is used: the {\em apocenter criterion} introduced by
Takahashi et al.\@ (1997).  In the apocenter criterion, a particle is
removed if its apocenter distance $r_{\rm a}$, which is a function of
$E$ and $J$, exceeds the tidal radius (see Fig.\,\ref{fig:Pot}).  The
energy criterion removes a larger number of stars than the apocenter
criterion.  For example: a star with energy equals to the tidal energy
cannot reach the tidal radius, except if its orbit is purely-radial,
i.e.; zero angular-momentum. This is illustrated in
Fig.\,\ref{fig:Pot}.


Both CW and Takahashi et al.\@ (1997), removed particles from the
cluster immediately after the escape criterion is satisfied.  This
assumption is justified if the orbital timescale at the tidal radius
is negligible compared with the relaxation time.  In real globular
clusters this is generally the case, but in the small $N$-limit where
$N$-body models operate this criterion is violated and stars are
usually removed from the stellar system too quickly.

Since a star has to move from one end of the cluster to the other, it
is important to account for the travel time of an escaping
star.  In our treatment an escaper timescale is introduced by
applying the following formalism for escapers (see Lee and Ostriker
1987, LO): 
\begin{equation}
\frac{df}{dt} = - \alpha_{\rm esc} f \left[ 1-\left(\frac{E}{E_{\rm
t}}\right)^3\right]^{1/2} \frac{1}{2\pi}\sqrt{\frac{4\pi}{3}G\rho_{\rm
t}}. 
\label{Eq:LO}\end{equation}
Here $E_{\rm t}$ is the tidal energy (the potential energy at the
tidal radius), $\rho_{\rm t}$ is the mean mass density within the
tidal radius, $G$ is the gravitational constant, and $\alpha_{\rm
esc}$ is a dimensionless constant which determines how quickly
escapers leave the cluster.  Note that there is a misprint (concerning
the factor $2\pi$) in their original equation (Eq.\,3.5) of LO.  A
star in an escaping orbit leaves the cluster within its orbital
timescale, which is --on average-- comparable to the crossing time for
the tidal radius.  The parameter $\alpha_{\rm esc}$ relates the
timescale on which escapers are removed from the cluster relative to
its dynamical timescale.  It is therefore expected that $\alpha_{\rm
esc}$ is of order unity.  We can determine its value by calibrating
the Fokker-Planck results to $N$-body results. The Coulomb logarithm
was taken as $\log \Lambda = \log N$.

Equation (\ref{Eq:LO}) is derived assuming the presence of the tidal
force for the escaping stars: $df/dt=0$ at $E=E_{\rm t}$.  Our model
computations include a tidal cutoff rather than a self consistent
tidal field and equation (\ref{Eq:LO}) is, strictly speaking, not
applicable. However, the most important improvement of equation
(\ref{Eq:LO}) is that escaping stars take time (of order of a crossing
time) before they are actually discarded from the cluster. In
principle, Eq.\,\ref{Eq:LO} could be modified also for anisotropic
models. However, we did not make any chances in Eq.\,\ref{Eq:LO}.

Stellar evolution in the Fokker-Planck computations is performed with
the same stellar evolution model as is used in the $N$-body
computations. For a better comparison with CW's Fokker-Planck
computations we performed for a few runs the same stellar evolution
treatment as they adopted.

\subsection{Initial conditions}

All clusters initially have the same half-mass relaxation time as in
the models IR of PZHMM, which is 2.87~Gyr.  The other conditions are
taken identical to that of CW's family 1.  The dimensionless depth of
the initial King model $W_\circ$ is 3. Mass function of the form
$dN(m) \propto m^{-2.5}$ between $0.4\solarm$ and $15\solarm$ is
used. All clusters initially fill their Roche-lobe; the King radius
equals the tidal radius. In the $N$-body model, stars that are outside
the tidal radius are removed.  This simple cutoff was chosen in order
to facilitate direct comparison with the Fokker-Planck results.

Apart from testing the various escape treatments in the Fokker-Planck 
models the only parameter which we change is 
the number of stars (see PZHMM for more details).

\section{Results}

\subsection{Comparison with Chernoff \& Weinberg}

\begin{figure}
\psfig{file=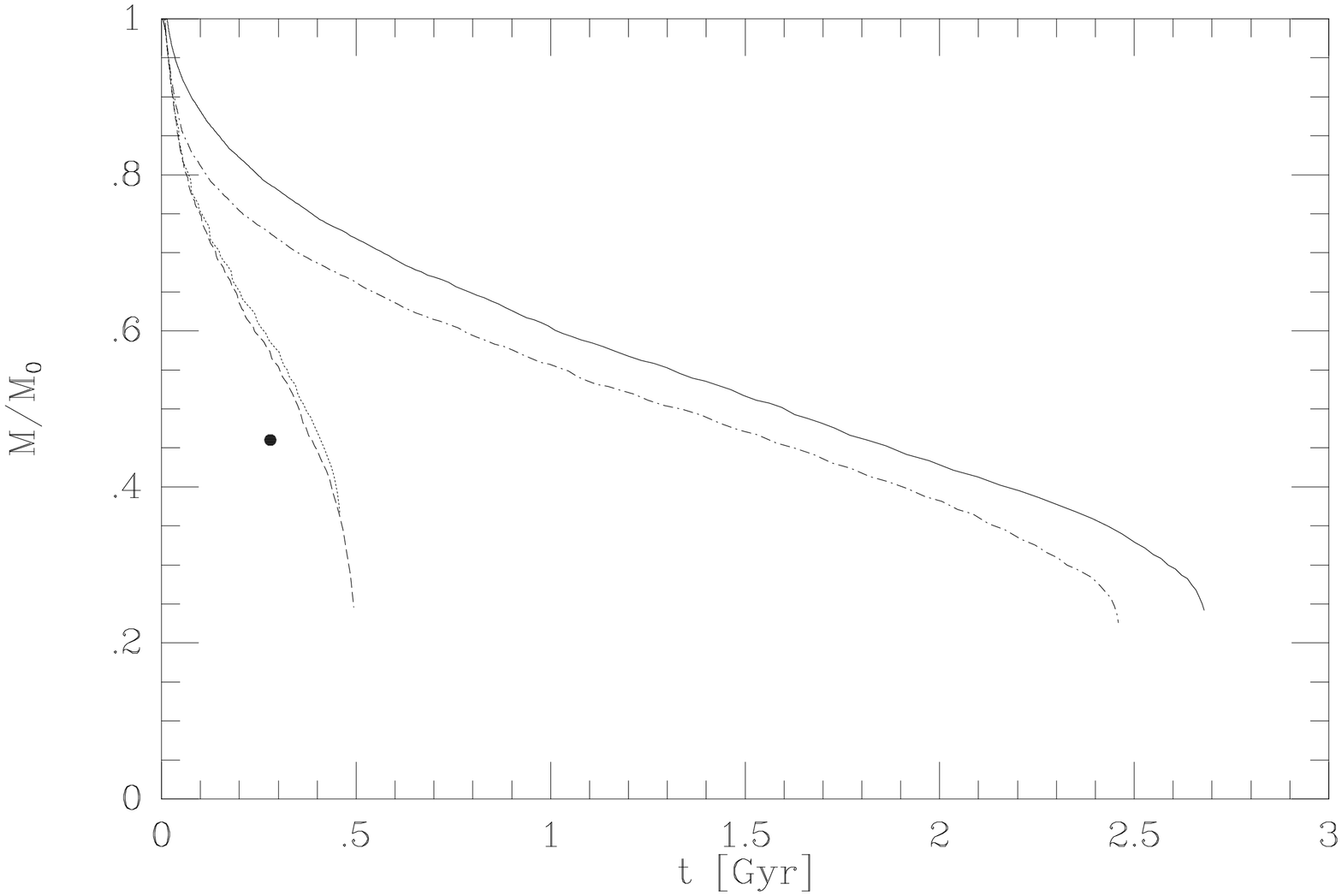,bbllx=540pt,bblly=30pt,bburx=70pt,bbury=730pt,width=10.0cm,angle=-90}
\caption[]{ The total mass of the simulated clusters (normalized to
the initial mass) as a function of time for different Fokker-Planck
models in which the number of particles is $\infty$ (by
definition). \\ The results of CW is presented as a $\bullet$ (to the
left) at the mass and age of the system where they considered the
cluster to cease to exist.  The models in which the energy criterion
is used are presented as the dashed line for the isotropic model {\em Ie}
and the dotted line for the anisotropic model {\em Ae}. \\ The {\em two}
lines to the right give the results of the anisotropic Fokker-Planck
model in which the apocenter criterion is used (model {\rm Aa}).  The
dash-dotted line uses the same stellar evolution model as is adopted
by CW and for the solid line the stellar evolution program \seba\ is
used. \\ All runs were stopped at the points where the self-consistent
potential could not be found.
\label{fig:CW}
}
\end{figure}

Figure\,\ref{fig:CW} shows the
evolution of the total mass of the star cluster (normalized to its
initial value) as a function of time.  The results of CW's computation
is presented as a dot in fig.\,\ref{fig:CW} (taken from their table
5). This is the end point of the simulation which CW regarded as the
end of cluster lifetime (disruption).

Our isotropic Fokker-Planck model (denoted as model {\em Ie}: {\em I} for
isotropic and {\em e} for the energy criterion) is given as a dashed line.
The same stellar evolution model and the same mass bins (20 mass bins)
as CW are used for this model. Therefore the result should coincide
with that of CW's corresponding run.  The agreement, however, is not
very good.  Our run reaches CW's end mass almost 40\% later.  We
repeated computations using several different sets of time steps and
numbers of mass bins, but the result did not change very much.  A
series of comparison runs with other initial conditions shows that
there is a tendency that the agreement improves for models with a
longer lifetime.  We did not investigate further the origins of this
disagreement, but rather decided to choose the result of the $N$-body
simulations as a base of our discussion.


A second run with the anisotropic Fokker-Planck model (denoted as {\em Ae},
where {\em A} stands for anisotropic) is
presented in fig.\,\ref{fig:CW} as a dotted line.  The difference
between the isotropic model ({\em Ie}) and the anisotropic model ({\em Ae}) is
small (see fig.\,\ref{fig:CW}). In both models the same stellar
evolution prescription as adopted by CW was used.

The largest difference is between models with the energy criterion and
the apocenter criterion (model {\em Aa}, where {\em a} stands for
apocenter criterion). The disruption time for model {\em Aa} is about
five times longer than that for the models {\em Ie} and {\em Ae}.  The
evolution of models {\em Ie} and {\em Ae} are similar when the ratio
of the tidal radius to the half-mass radius is small (Takahashi and
Lee 1998, in preparation). This is because a strong tidal cutoff (as
in a King model) suppresses the development of anisotropy in the halo. 
The apocenter criterion allows particles which would have escaped
while using the energy criterion to stay in the cluster.  The escape
rate in models which use the apocenter criterion is therefore
considerably slower than in models which use the energy criterion (see
Fig.\,\ref{fig:Pot}).

\subsection{Effects of stellar evolution models}

The stellar evolution model used by PZHMM is different from that
adopted by CW.  In the computations of CW the post main-sequence
evolution of the stars is neglected and stars in PZHMM's model live
therefore somewhat longer. In fig.\,\ref{fig:CW} the results of two
models {\em Aa} are presented of which one is computed using the
stellar evolution model of CW (dash-dotted line) and the other of
PZHMM's model (solid line).  The difference in the evolution of the
mass of the star clusters is small, as excepted.  The dissipation time
of the two models differ by less than 10\%.

\subsection{$N$-dependence of the cluster evolution}

For all computations in this section, we use the stellar evolution
models according to the prescription in \seba\
and employ Eq.\,\ref{Eq:LO} as escape condition.

\begin{figure}
\psfig{file=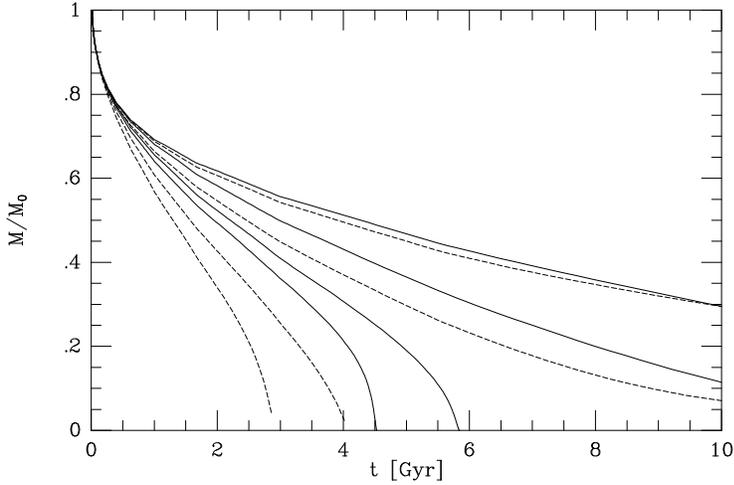,bbllx=540pt,bblly=30pt,bburx=70pt,bbury=730pt,width=10.0cm,angle=-90}
\caption[]{ Mass as a
function of time for a number of Fokker-Planck models.
The four solid lines represent the results of model
Aa with 32k, 16k, 4k and 1k particles from left to right,
respectively.  Dotted curves present model {\em Ie} for the same numbers of
particles as for model {\em Aa}. \\
The time scale for escapers via Eq.\,\ref{Eq:LO} 
with $\alpha_{\rm esc} = 2$ for all models.
\label{fig:Aa}
}
\end{figure}

Figure\,\ref{fig:Aa} shows the results of calculations with
$\alpha_{\rm esc}=2$ (see Eq.\,\ref{Eq:LO}) in models {\em Ie} and {\em Aa}.  The
choice for $N$ at which we should calibrate $\alpha_{\rm esc}$ is
rather delicate. The results of the Fokker-Planck computation is more
sensitive to $\alpha_{\rm esc}$ for small $N$ than for large $N$.
However, for a smaller number of particles the $N$-body results tend
to become more noisy. We decided to use a modestly large number of
stars ($N$ = 16384, 16k) to calibrate $\alpha_{\rm esc}$.  It turns out
that $\alpha_{\rm esc} =2$ gives the best agreement.

\begin{figure}
\psfig{file=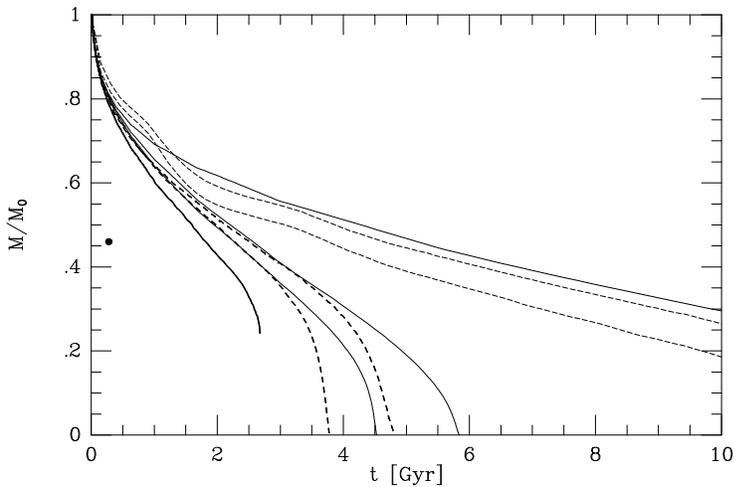,bbllx=540pt,bblly=30pt,bburx=70pt,bbury=730pt,width=10.0cm,angle=-90}
\caption[]{ Mass as a function of time for the $N$-body models
(dotted lines) and Fokker-Planck models {\em Aa} (solid lines, see also
Figs.\,\ref{fig:CW} and Fig.\,\ref{fig:Aa}).  The {\bf thick} solid
line to the left is for $\infty$ stars, the three subsequent solids
are for 32k, 16k and 1k stars, respectively. The left and right thick
dashed lines give the results of the $N$-body simulations for 32k and
16k stars, respectively. The two {\sl thin} dashed lines represent the
${1 \over 2} \sigma$ deviation from the mean of the 10 performed runs
with 1k stars. \\ The results obtained by CW is presented as a
$\bullet$.
\label{fig:Nb}
}
\end{figure}

Figure\,\ref{fig:Nb} presents the results of a number of $N$-body
computations and compares these with the results of the Fokker-Planck
models {\em Aa} with $\alpha_{\rm esc}=2$.  In order to minimize the
statistical fluctuations in the $N$-body results we performed 10
identical computations with $N$ = 1024 (1k). For economic reasons we
performed only three runs with $N$ = 16k and a single run with 32k
stars. Each of the 1k runs took about an order of magnitude less
computer time on GRAPE-4 than one of the anisotropic Fokker-Planck
computations on a fast workstation (the $N$-body with 32k stars took
approximately two orders of magnitude longer than the Fokker-Planck
models, i.e.: almost three weeks).  However, even with the mean of 10
runs the noise in these 1k computations is rather large (see
Fig.\,\ref{fig:Nb}).  The $N$-body computation with $N=32$k is, due to
historical reasons, performed with an upper mass limit of 14\,\msun\
instead of 15\,\msun.  The lifetime of this model is therefore
expected to be slightly longer than if 15\,\msun would have been used. 
However, the difference is small, which we tested by using different
mass cut-offs in the Fokker-Planck model.


The agreement between the Fokker-Planck results and the $N$-body model
is quite good although there are still some deviations. 
After about 70\% of the mass is lost, the deviation becomes noticeable.
This may be related to the disruption of the cluster on the dynamical
time scale as discussed by CW, Fukushige and Heggie (1995) and
PZHMM. Another effect, which is most clearly visible in the $N$-body
model with fewest particles, is the dip of the mass after about
a billion years.

\section{Conclusions}
We have found the reason why the Fokker-Planck calculations of CW and
the $N$-body calculations of Fukushige \& Heggie (1995) and PZHMM gave
very different results. The assumption of velocity isotropy and the
over-simplified escape criterion (the energy condition and removing
stars instantaneously) caused an enormous overestimate of the escape
rate. By using an anisotropic Fokker-Planck model with an improved
escape criterion, we have succeeded to achieve excellent agreement
between Fokker-Planck and $N$-body results.

The dependence of the dissipation time on the number of particles is
also understood. Stars need some time to travel away from the cluster
in order to be gobbled up by the galaxy.  This timescale is of the
order of a crossing time at the tidal radius.  Therefore the escape
rate depends on the ratio of the relaxation time to the dynamical
time, i.e.; on the number of stars.

\acknowledgments We are grateful to Toshiyuki Fukushige, Douglas
Heggie (referee), Piet Hut, Junichiro Makino and Steve McMillan for many
discussions and software development. SPZ thanks Atsushi Kawai for
keeping GRAPE in shape while performing the computations. This work is
supported in part by the Research for the Future Program of Japan
Society for the Promotion of Science (JSPS-RFTP97P01102).

\clearpage

\end{document}